# Design and discovery of a novel Half-Heusler transparent hole conductor made of all-metallic heavy elements

F. Yan[1,†], X. Zhang[2,†], Yonggang Yu[2], L. Yu[2], A. Nagaraja[1], T. O. Mason[1], and Alex Zunger[2,*]

[1]Department of Materials Science and Engineering, Northwestern University, Evanston, IL, 60208, USA

[2]University of Colorado, Boulder, CO 80309, USA

[†] Equally contributing authors

**Metallic conductors that are optically transparent represent a rare breed of generally contraindicated physical properties that are nevertheless critically needed for application where both functionalities are crucial. Such rare materials have traditionally been searched in the general chemical neighborhood of compounds containing *metal oxides,* expected to be wide gap insulators that might be doped to induce conductivity. Focusing on the family of 18 valence electron ABX compounds we have searched theoretically for the ability of the compound's electronic structure to simultaneously lead to optical transparency, in parallel with the ability of its intrinsic defect structures to produce uncompensated free holes. This led to the prediction of a stable, never before synthesized TaIrGe compound made of *all-metal heavy atom compound* as the 'best of class' from the V-IX-IV group. Laboratory synthesis then found it to be stable in the predicted crystal structure and *p*-type transparent conductor with measured strong direct absorption of 3.36 eV**

**and remarkably high (albeit not predicted) hole mobility of 2730 cm$^2$/Vs at room temperature. This methodology opens the way to future searches of transparent conductors in unexpected chemical groups.**

*alex.zunger@colorado.edu

*Design principles needed for searching compounds with contraindicated physical properties:* Transparent conductors (TC's) needed in solar cell, light emitting diode and flat panel displays[1-5], represent the usually contraindicated functionalities of optical transparency (generally associated with electrical insulators) coexisting with electrical conductivity (generally associated with optically opaque metals)[6]. The search of TC's has generally focused on chemical groups characterized as *light element wide gap oxides,* such as Al-doped ZnO,[7] (for *n*-type) or CuAlO$_2$,[8] and K-doped SrCu$_2$O$_2$,[9] (for *p*-type) TC's. Rather than attempt discovery within such a preconceived chemical neighborhood (either by high-throughput computation[10] or via combinatorial synthesis), the discovery of TC's might be guided instead by a set of physics based metrics that *p*-type TC's need to satisfy *a priory* [9,11-14] then searching not only light element oxides or nitrides but allowing for a broader range of chemistries to be inspected. These "Design Principles" include a combination of (a) electronic structure features of the perfect crystal and, in parallel, (b) specialized properties of the non-ideal defected lattice. In category (a) we require a wide (> 2.5 eV) direct band gap for optical transparency (but the non-absorbing indirect band gap can be significantly smaller), whereas good mobility requires rather light (< 0.5 $m_0$) hole effective mass and a bulk hole wavefunction that avoids as much as possible the close neighborhood of the ions. In category (b) we require that the intrinsic defects that are hole-producers (acceptors such as cation vacancies) have low formation enthalpy (i.e., be abundant) and have shallow acceptor levels (i.e., be readily ionizeable), whereas the intrinsic defects that are 'hole-killers' (donors such as anion vacancies) must have high formation enthalpy or have levels that are electrically inactive (not ionizeable). In considering such design principles it is not obvious *a-priori* that only the traditionally sought light element oxides are eligible for satisfying these simultaneous conditions.

***Known as well as 'missing' compounds in the ABX family***: We chose to focus on the 18-electron ABX compounds that represent diverse chemical groups such as $A^{n}B^{10}X^{(8-n)}$ as well as $A^{(n+1)}B^9X^{(8-n)}$ ($n$ = 1, 2, 3, 4) with atoms spanning columns 1~17 in the Periodic Table. This broad range of atoms (albeit within the group of ABX) encompasses a broad range of chemistries including light as well as heavy elements; anions as well as cations. The previously synthesized members from these groups manifest extraordinary functionalities such as thermoelectricity, superconductivity, piezoelectricity, and topological insulation, but as yet no TC's. But not all atom combinations that by the current understanding of solid-state chemistry can plausibly lead to such ABX structures were in fact realized. For example, from a total of 483 possible 18 valence electron ABX compounds from the above noted chemical groups, only 83 are known,[15] whereas 400 are, in fact, "missing compounds" that might constitute an attractive playing ground for new materials with new functionalities. We[16, 17] and others[18,19] have developed theoretical, first-principles techniques for examining the thermodynamic stability of missing compounds sorting out the 'missing and predicted unstable' from the 'missing and predicted stable'. Our first principles thermodynamics search includes the examination of (a) The lowest-energy crystalline form of the ABX phase, (b) its stability with respect to decomposition into any combination of elemental or binary phases of the A+B+X constituents and (c) the dynamic (phonon) stability of the lowest energy ternary phase. Note that in search (a) we do not limit the allowed structure to just a single structure-type (as in Carrete *et al*[18], where only the *cubic* LiAlSi-type structure was considered, leading to the elimination of candidate ABX compounds that are stable in other structure types, or in Ref. 19 where only the *cubic perovskite* structural form of $ABO_3$ were considered for electrocatalysis. Note further that unlike Refs. 18, 19 in (b) we systematically correct the density functional values of formation energies[16], thus avoiding false elimination of stable ternaries on account of the underestimation of their enthalpies. We note that our search is different than Ref. 10 where only previously documented oxide materials were examined. A succinct summary of the method we use is given in Supplementary Section I.

The application of first principles thermodynamics based on density-functional methodology[16,17,20] to a total of 483 ABX compounds (400 missing as well as 83 known[15]), we have predicted that the 54 of the 400 missing compounds are predicted stable in specific crystal structures and 346 are predicted unstable. As a test to our thermodynamic equilibrium methodology we applied it to ~30 known (previously synthesized and structurally characterized) ABX compounds listed in ICSD[15]. We find in all cases that these compounds are predicted to be stable in the *observed* crystal structures, validating the methodology. In the present work we do not focus on the broad picture of the prediction of stable new compounds[20]. Instead we wish to focus on the functionality of transparent conductivity. Figure 1**a** provides the calculated results on one such group V-IX-IV of 18-electron ABX compounds, with check marks indicating previously known compounds, plus signs (+) indicating previously missing and now predicted stable, and minus signs (-) indicate previously missing and now predicted unstable compounds. We next illustrate how a specific functionality is searched.

***Design-principle guided search: (a) Properties of ideal bulk crystal.*** Having supplemented the currently known groups of 18-electron ABX compounds by the previously missing and now predicted stable compounds, we next examine their systematic electronic properties using self-consistent hybrid functional (HSE06)[21] wavefunctions considering the design metrics for *p*-type transparent conductors. We find a number of new structure-property relationships in this family that focused our attention to specific subgroups for prediction *p*-type transparent conductors.

*First*, we find that all 18-electron cubic Half-Heusler ABX compounds containing *two* transition metal atoms (e.g., TaIrSn and ZrIrSb) are non-metals (i.e., have a band gap between occupied and unoccupied bands), whereas the cubic 18-electron Half-Heusler compounds with one transition metal atom (e.g., AlNiP) are metals (see Supplementary Figs. S3 and S4). Thus, to look for candidate *transparent* materials we focus on the former group containing two transition elements. The finite band gap in the former case is explained by the strong *d-d* hybridization between the two transition metals (e.g. Ta and Ir) that are nearest neighbors in the Half-Heusler (LiAlSi-type) structure, as illustrated for TaIrGe in Fig. 1**c**.

*Second*, we find that the ABX compounds with *heavy X atom tend to be* stable in cubic Half-Heusler structures, hence can potentially have wide band gaps. For example, the predicted ABX compounds in groups IV-X-IV, IV-IX-V, and V-IX-IV with heavy X elements Sb, Bi, Sn, or Pb are all stable in cubic structure. In contrast, we find that ABX with light X atoms (i.e. O, S, Se, N, P, As, C, Si, and Ge) tends to have non-cubic structure and thus are often metallic, since for light X (e.g. O), the A and B metals are ionized and strongly repel each other when they are nearest neighbors as in the LiAlSi-type structure. We thus look for TC's in heavy X atom compounds.

In accordance with the above two structure-property relationships, we identify nine wide gap ($E_g^{dir} \gtrsim 2.3$ eV) members in the groups of cubic, two-transition element Half-Heusler ABX compounds: from V-IX-IV group we find TiCoSb, TiRhSb, TiIrSb, ZrRhSb, ZrIrSb and HfCoSb; from the IV-X-IV group we find HfPtSn, whereas from the V-IX-IV group we identify TaIrGe and TaIrSn. These are *heavy, all-metal atom Half-Heusler insulators,* rather unusual in that many heavy atom compounds tend to be narrow gap semiconductors[22] or semimetals[23] (e.g. HgTe, PbSe, PbTe, $Bi_2Se_3$, and InSb). The fundamental band gaps of the above nine Half-Heuslers are indirect (as illustrated for TaIrGe in Fig. 2**a**). Among them, TiIrSb, ZrIrSb, TaIrGe and TaIrSn are newly predicted never before synthesized compounds (see Supplementary Section I).

**Results on Properties of ideal bulk crystals: Thermodynamic stability and crystal structure of the heavy, all-metal atom insulators:** The ranges of

thermodynamic stability are evaluated in the chemical potential space of A, B, and X considering the elemental, binary and ternary competing phases (see the inset of Fig. 3**a** and Supplementary Section II). The predicted allowed chemical potential range reflects the thermodynamic growth conditions, which can be used as a guide for experimental synthesis. For the newly predicted and never before synthesized ABX compounds (TiIrSb, ZrIrSb, TaIrGe and TaIrSn), we find that their stability range is located mainly around the boundary of X richest and A/B-poor condition.

The predicted lowest-energy structure-type of TaIrGe is shown in Fig. 1**b**. In this structure, both A and X atoms have $B_4$ tetrahedrally coordinated environments, and the B atoms are eightfold coordinated by four A atoms and four B atoms. There is an empty interstitial site as shown by the small black sphere in Fig. 1**b**, having the same local environment (i.e. $A_4$ and $X_4$ tetrahedrons) as B atoms. Next we focus on TaIrGe since it is the widest direct gap material and the only cubic Germanide in the 18-electron ABX family (see Supplementary Section I).

***Results on Properties of ideal bulk crystals: Electronic structure and optical transparency:*** In the cubic Half-Heusler structure of TaIrGe, all the atomic sites in TaIrGe have $T_d$ symmetry, so the *p* and *d* orbitals occur in the $T_2$ and *E* representations of the $T_d$ point group. As illustrated in Fig. 1**c** (see also calculated density of states in Fig. 2**b**), the $T_2(Ta, d)$ orbital strongly couples with $T_2(Ir, d)$, and $E(Ta, d)$ couples with $E(Ir, d)$, opening a large gap between anti-bonding and bonding states. According to our HSE06+GW electronic calculation (see Supplementary Section I), the compound has an indirect fundamental band gap of 1.74 eV. The valence band maximum (VBM) and conduction band minimum (CBM) are at L valley and X valley, respectively, and the smallest energy gap between VB and CB (i.e. direct bandgap, 2.64 eV) is at X valley (see Fig. 2**a**). The calculated optical absorption coefficient of TaIrGe based on the GW approximation[24] for electron's self-energy is presented in Fig. 2**c** showing an optical transition onset of ~2.64 eV. Thus TaIrGe is transparent for most visible light with frequency below 2.64 eV. The first strong peak of absorption coefficient appears at 3.1 eV corresponding to the transition between the two parallel bands highlighted by green in Fig. 2**a**. The absorption coefficient is < $10^5$ cm$^{-1}$ for photon energy in the range of (2.64 eV, 2.9 eV) which is acceptable for thin film (a few hundred nm) transparent conductors.

**_Results on Properties of ideal bulk crystals: The nature of the hole orbitals promises ideal carrier mobility._** Three factors are noteworthy:

*First*, the hole wavefunction (see Fig. 2**d**) is delocalized mainly along the *interstitial channels* (in between Ta-Ir and Ta-Ge atoms) that do not pass through ionic sites and hence have but a low potential to scatter the mobile carriers.

*Second*, there are quite a few hole states in close energetic proximity to each other, so the degeneracy of the hole states in the three-dimensional momentum space is rather high: the hole Γ valley (3-fold band degeneracy) and the 6-fold W valley (single band) lie just below the 4-fold L valley (2-fold band degeneracy).

*Third*, the predicted low electron and hole effective masses: It is generally difficult to find oxides with not so high hole masses because the valence band tends to be narrow and leading to large effective masses[10]. Here, we calculate the electron and hole effective masses for TaIrGe, which has no oxygen, and excluding the oxygen localization. The achieved electron effective masses (near CBM at *X* point) are $m^*_{e,\parallel} = 0.54$ and $m^*_{e,\perp} = 0.27$, and the heavy hole (HH) and light hole (LH) effective masses (near VBM at *L* point) are $m^*_{HH,\parallel} = 0.96$, $m^*_{HH,\perp} = 0.77$, $m^*_{LH,\parallel} = 0.96$ and $m^*_{LH,\perp} = 0.39$, respectively. All the hole effective masses are significantly lower than that of the conventional *p*-type transparent *oxides*[10], such as $CuAlO_2$ and $SrCu_2O_2$, which have hole effective masses larger than 2.

***Design-principle guided search: (b) Defect property and p-typeness***. Having isolated an interesting candidate TC material with acceptably large band gaps and consequent *optical transparency*, we next follow the crucial design principle for *p*-typeness: promote hole-producers but suppress hole-killers. Our first-principles calculations focused on two key quantities for defects: (i) the defect formation energy $\Delta H$ ($q$, $E_F$, $\{\mu\}$), which is a function of the defect charge state $q$, the Fermi-energy $E_F$, and the chemical potential $\{\mu\}$ of atoms involved in forming the defect (reflecting the growth conditions), and (ii) the defect charge transition energy (donor or acceptor levels), $\varepsilon(q/q')$, defined as the $E_F$ where the $\Delta H$ of a defect at two different charge states $q$ and $q'$ are equal. Taking into account $\Delta H$ for all intrinsic defects, one can determine the thermodynamic equilibrium Fermi energy $E_F^{(eq)}$ (see Figs. 3 & 4**a**) and the carrier concentration predicted at a given temperature and chemical potential (Fig. 4**b**).

***Results on Properties of doped p-type crystals: Hole producers***. A most exciting prediction from our intrinsic defect simulation is that TaIrGe displays a strong *p*-type behavior ($n_{hole}$ > $10^{16}$ cm$^{-3}$, $E_F^{(eq)}$ < $E_{VBM}$ + 0.2 eV) in the majority of its stability area (bright blue color in Fig. 4**a** and bright red color in Fig. 4**b**). From the simulation for crystal growth condition (compounding at 1200 K followed by a quench to room temperature), we find the *maximum* hole concentration of ~2.5×$10^{17}$ cm$^{-3}$ when approaching the P$_{max}$ corner of the stability area (the most Ta-poor and Ge-rich condition in Fig. 4**b**). The hole concentration could be further increased by introducing extrinsic dopants into TaIrGe.

The microscopic origin for *p*-type is revealed by inspecting formation energies of leading defects under the optimal chemical potential conditions (inset in Fig. 3**a**), P$_{max}$ (conducive to holes [p-type]) and N$_{max}$ (conducive to electrons [n-type]). Out of the twelve intrinsic point defects, including three vacancies ($V_{Ta}$, $V_{Ir}$, $V_{Ge}$), three interstitials ($Ta_i$, $Ir_i$, $Ge_i$), and six antisite defects ($Ta_{Ir}$, $Ir_{Ge}$, $Ir_{Ge}$, $Ge_{Ir}$, $Ta_{Ge}$, $Ge_{Ta}$), we identified Ge-on-Ta antisite defect in (1-) charge state as the dominant acceptor, having lower formation energy than other defects at the Ta-poor conditions (P$_{max}$, Fig. 3**a**). This acceptor state is derived from the unfilled shallow Ge-*4p* state at the Ta-site forming chemical bonds with its nearest neighbor Ir *6p* and *5d* orbitals. The neutral to (1−) charge state transition in Ge$_{Ta}$ occurs at $E_v$+0.21 eV. Considering the 1.74 eV band gap, this transition level is rather close to VBM. The holes produced by Ge$_{Ta}^{1-}$ are not compensated by potential hole-killer defects (i.e. metal interstitials), that have much higher formation energies (Fig. 3**a**).

***Laboratory Synthesis and crystallographic structure determination:***
Experimental realization of the hitherto missing (unreported) TaIrGe was accomplished by bulk synthesis methods, followed by optical and electronic characterization. The TaIrGe specimens were synthesized by vacuum annealing the stoichiometric mixture of pure elements (see details in Supplementary Section III). Figure 5**a** shows the X-ray diffraction pattern of TaIrGe, confirming that phase-pure bulk TaIrGe was successfully fabricated. By comparing the experimental X-ray pattern with the calculated diffraction pattern, positions and intensities of the experimental TaIrGe are in agreement with the theoretical prediction, and the refined lattice parameter is 5.9664(5) Å with a calculated density of 13.94 g/cm$^3$. The lattice parameters were accurately established using high purity silicon as an internal standard. Figure 5**b** displays a representative bright field TEM image and corresponding selected area electron diffraction (SAED) pattern, suggesting a grain size of around 100 nm, and with no secondary phase.

***Measured optical properties:*** The optical absorption spectrum of bulk TaIrGe, obtained from the ultraviolet-visible diffuse reflectance measurements (see Fig. 6**a** and Supplementary Section III) indicated that TaIrGe exhibits two optical absorption thresholds at about 365 and 485 nm in the visible range (which correspond to band gap energies of 3.39 and 2.55 eV, respectively). Subsequent analyses of the diffuse reflectance using the Kubelka-Munk function[25-27] indicated an indirect transition of 1.64 eV (Fig. 6**b**) and a direct transition of 3.36 eV (Fig. 6**c**). The difference between measured optical band gaps and predicted values (~3.1 eV) is ~0.2 eV. Thus, the absorption window of TaIrGe covers most of the visible solar spectrum (~ 350 to 700 nm). To ascertain about this optical band gap, we performed high vacuum thin film deposition using TaIrGe target in a pulsed laser deposition system and the obtained 30-nm-thick film shows transparent in visible light. Figure 6**d** shows the optical transmittance spectrum of the TaIrGe film grown on transparent quartz substrate, which varies from 85% to 92% in the wavelength region of 350 to 700 nm. This demonstrates that the TaIrGe film is transparent in the visible region.

***Measured electrical transport properties:*** The electrical conductivity ($\sigma$) of the bulk specimen, measured using Van der Pauw method at room temperature, was ~ 0.35 S/cm. The Hall coefficient ($R_H$), measured using a five-probe configuration, gave +7.8×10$^3$ cm$^3$/C, indicative of *p*-type conduction, which is in agreement with the results from the intrinsic defect calculation (A Seebeck coefficient was found to be +82 $\mu$V/K at room temperature for the TaIrGe bulk, further confirming that the TaIrGe is a *p*-type semiconductor). Combining the conductivity and Hall coefficient resulted in a hole concentration of ~0.8× 10$^{15}$ cm$^{-3}$ and a Hall mobility of ~+2730 cm$^2$/Vs. The achieved Hall mobility is much higher than that of known *p*-type transparent conducting oxides[8,9](e.g. ~10 cm$^2$/Vs for CuAlO$_2$).

**Discussion**

We illustrate that a rather rare functionality of *p*-type transparent conductivity can be discovered by inverse design in a never before made compound. Whereas this functionality has been traditionally searched in the general chemical neighborhood of compounds containing *light element oxides,* we use a different search strategy that led to the identification of a good *p*-type TC in an *all-metal heavy atom compound* TaIrGe. A remarkable high hole mobility 2730 cm$^2$/Vs has been found in TaIrGe by experiment at room temperature. This study opens the way to search transparent conductors among non-oxide heavy-element materials and to designing high mobility transparent electronic devices. Some of the important insights gained include:

*(i)* Filling the gaps of 'missing compounds' in a group of materials allowed a confident formulation of interesting structure-property chemical rules enabling zooming effectively to the subgroup likely to contain *p*-type TC's. These include the rules that the 18-electron ABX with heavy X atoms tends to have cubic Half-Heusler structure and that insulation band gap are exclusively correlated with cubic Half-Heusler structure having with two transition metals.

*(ii)* Defect physics in the TaIrGe 18-electron ABX Half-Heusler prototype: The X$^{IV}$-on A$^V$ antisite *hole-producer* has low formation energy since both A and X sites are in a similar four-fold coordination (while B in eight-fold) and the size of X$^{IV}$ (Ge) is smaller than A$^V$ (Ta). In contrast to the favorable conditions for producing holes, the *hole-compensating* (electron producing) metal interstitials (Ir$_i$, Ta$_i$, Ge$_i$) that form by occupying the interstitial site in Half-Heusler structure (see Fig. 1**b**), require high formation energy due to the large atomic sizes (for Ir$_i$ and Ta$_i$) or repulsion from iso-atom nearest neighbors (for Ta$_i$ and Ge$_i$) and thus do not block the acceptors.

*(iii) Electronic structure conducive to high hole mobility:* The previously studied *p*-type transparent conductors are prone to have low hole mobilities[8,9] (e.g. ~10

cm$^2$/Vs for CuAlO$_2$ and 0.46 cm$^2$/Vs for SrCu$_2$O$_2$), which limited their application. In contrast, the Half-Heusler compound TaIrGe can have a hole mobility as high as the standard *covalent* semiconductor Ge (~1900 cm$^2$/Vs at room temperature[28]). The electronic structure features that enable good hole mobility include the distribution of the hole VBM wavefunction along interstitial channels that do not pass through ionic sites; the high degeneracy of the hole states in the 3D momentum space near VBM, and the relatively low hole effective mass due to significant hybridization (Fig. 2d). Thus, whereas TaIrGe has hole property similar to those of narrow gap semiconductors such as Ge, unlike the latter cases it has a wide band gap opened by the strong *d-d* hybridization between the metal atoms.

 *(iv) Inverse design of functional materials:* The proposed approach in inverse design framework *starts* from the physically formulated design principles of material functionalities, without the preconceived limitations referring to the previous discoveries, *considers* both existing compounds as well as previously undocumented but now predicted stable members followed by evaluation of design metrics, thereby (significantly) increasing the playing field of discovery, and *leads* to the efficient selection of candidate materials for experimental realization. This approach, searching directly for specific target functionality (rather that making all compounds at the outset, as in combinatorial synthesis or its computational analogue—high throughput calculations), as demonstrated in current study, can find interesting candidate materials in unexpected chemical groups, of technologically significant functionalities that are hard to find and optimize by rational experiments (e.g. *p*-type TC), potentially opening a new direction to design functional materials (e.g. all-heavy-element materials for high-mobility TC). Following this approach, one can also start the design from the physical formulation of design principles, thus eventually leads to discovery of new material functionalities from inverse design.

**Methods**

**Thermodynamic stability calculations**. To determine the thermodynamics stability of the ABX compounds, the $\Delta H_f$ of each ABX compound and all possible competing phases were calculated. We use fitted elemental reference energies[16] to correct the DFT error on formation enthalpies ($\Delta H_f$). A compound is considered stable under thermodynamic equilibrium conditions if the values of the chemical potential are such that the formation of a given hypothetical ABX compound is energetically the most favorable of all possible competing phases[17] (See Supplementary Information II).

**Optical absorption calculations.** The optical absorption coefficient was calculated using the GW approximation perturbatively on the top of the wavefunctions and energy eigenvalues calculated from a generalized Kohn-Sham scheme with the hybrid exchange-correlation functional (HSE06)[21] (i.e. HSE06+GW see details in Supplementary Section I). The spin orbit coupling (SOC) effect in TaIrGe is non-negligible. We calculate the density of states from HSE06 with SOC to compare with that from HSE06 without SOC. The fundamental band gap of TaIrGe is slightly reduced (by less than 0.1 eV) by spin-orbit coupling (see Supplementary Section I).

The electron and hole effective masses are obtained by fitting the actual *E-k* diagram around CBM and VBM of energy eigenvalues calculated from HSE06, respectively.

**Defect calculation.** Defects are calculated in a 2×2×2 cubic supercell with 96 atoms using the same GGA+U functional as in the thermodynamic stability calculations. The band gap calculated by GGA+U is 1.24 eV, smaller than the GW value 1.74 eV. It is found that from GGA+U to GW, the VBM remains unchanged and CBM undergoes an uplift of 0.5 eV. To correct the finite supercell error, we applied the image charge corrections and band filling corrections as in Ref. 29. A total of 12 types of intrinsic point defects have been calculated, including three vacancies, three interstitial, and six antisite defects. We apply the tetrahedral bending mode to the nearest-neighbor atoms[30].

**Materials synthesis**. Polycrystalline TaIrGe samples were synthesized by vacuum annealing a mixture of pure elements in the ratio of 1:1:1 at ~1200 K for 2 days and quench to room temperature. (Details in the Supplementary Information).

**X-ray powder diffraction**. The phase structure of TaIrGe pellets was characterized using a Scintag XDS2000 diffractometer with a Cu-K$\alpha$ radiation source of wavelength 1.5406 nm operated at 40 kV and 20 mA and a liquid nitrogen-cooled Ge detector (GLP-10195/07-S) over 2$\theta$ angles in the range 20 ~ 80º. The lattice parameters were determined using an internal Si standard. The lattice parameters were refined based on whole pattern fitting (WPF) Rietveld refinement in the JADE 9.0 software package (see Supplementary Section III).

**Energy dispersive X-ray spectroscopy**. The chemical composition of the resulting samples was analyzed by Energy-dispersive X-ray spectroscopy (EDS) attached to a Hitachi Field emission scanning electron microscope (FE-SEM SU8030). Data was acquired by an accelerating voltage of 20 kV and a 1 min accumulation time.

**Transmission electron microscope**: TEM images and SAED pattern were collected on a Hitachi HD-2300A scanning Transmission Electron Microscope.

**Optical measurements.** The ultraviolet-visible diffuse reflectance spectra and transmittance spectrum of the samples were characterized at room temperature in air by a Perkin-Elmer Lambda 1050 UV-Vis/NIR spectrophotometer. Reflectance measurement of the bulk sample was recorded relative to a calibrated standard and the reference material was used to establish a baseline. The diffuse reflectance data were collected in the wavelength range of 250 ~ 1500 nm. Absorption spectra were calculated from the reflectance spectra by means of the standard Kubelka-Munk conversion. The optical band gap was determined from the associated Tauc plots[31]. The transmittance spectrum was determined at room temperature for both the substrate and film.

**Electrical transport measurement**. Four-probe resistivity was determined using the Van der Pauw method on the pellet samples by a Physical Measurement System (Quantum Design PPMS 6000). The Seebeck coefficient was obtained from the slope of the thermovoltage versus temperature gradient. The Hall coefficient ($R_H$) was measured at room temperature using a five-probe configuration with the magnetic sweep between ±2 Tesla using the PPMS system.

**Acknowledgements**

This work is supported by the U.S. Department of Energy, Office of Science, Basic Energy Sciences, under Contract No. DE-AC36-08GO28308 to NREL as a part of the DOE Energy Frontier Research Center "Center for Inverse Design". XRD patterns were collected and refined at the J. B. Cohen X-Ray Diffraction Facility supported in part by the National Science Foundation (DMR-1121262 MRSEC) at the Materials Research Center of Northwestern University. This work made use of the EPIC facility (NUANCE Center-Northwestern University), which has received support from the NSF-MRSEC (DMR-0520513), NSF-NSEC (EEC-0118025|003).

**Author Contributions**

F. Y. fabricated pure TaIrGe samples, carried out structure analysis by XRD and TEM, and performed electrical and optical property characterization. A. N. fabricated TaIrGe using the microwave-sintering method. X. Z. carried out theoretical calculations of bulk 18-electron ABX compounds; L. Y. performed the phonon and optical absorption spectra calculation; Y. Y., L. Y. and X. Z. performed the defect calculation and analysis. The paper was written by F. Y., X. Z., and Y. Y. with the help from L. Y. Here T. M. supervised the experimental work. A. Z. supervised all the theoretical studies, the analysis of the results and directed the writing of the paper.

**Additional information**

Supplementary information is available in the online version of the paper.

**Competing financial interests**

The authors declare no competing financial interests.

**Figures**

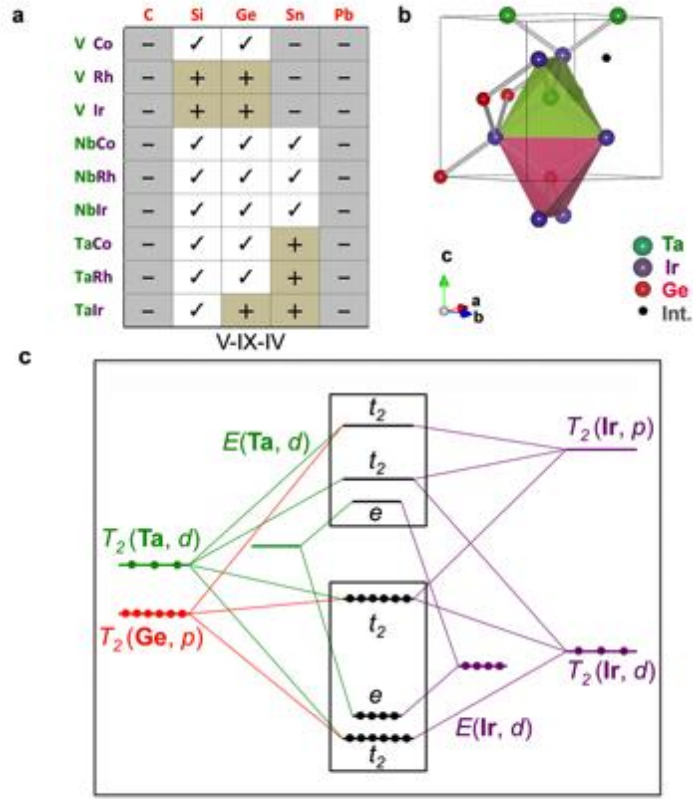

**Figure 1 | Crystal structure and orbital diagram of TaIrGe. a.** ABX compounds in V-IX-IV group. The compounds labeled by check marks (✓) are documented, the compounds marked by plus (+) are predicted stable, and the ones marked by minus (−) are unstable. **b,** The stable crystal structure (LiAlSi-type, space group: F-43m) of TaIrGe. The small black sphere (indexed as Int.) indicates the empty interstitial site that is equivalent to the Ir site. Two polyhedrons are shown to illustrate the similarity of first shell coordination environment (Ir$_4$ tetrahedron) of Ta and Ge. **c,** Orbital diagram of TaIrGe as an example of 18-electron ABX Half-Heuslers.

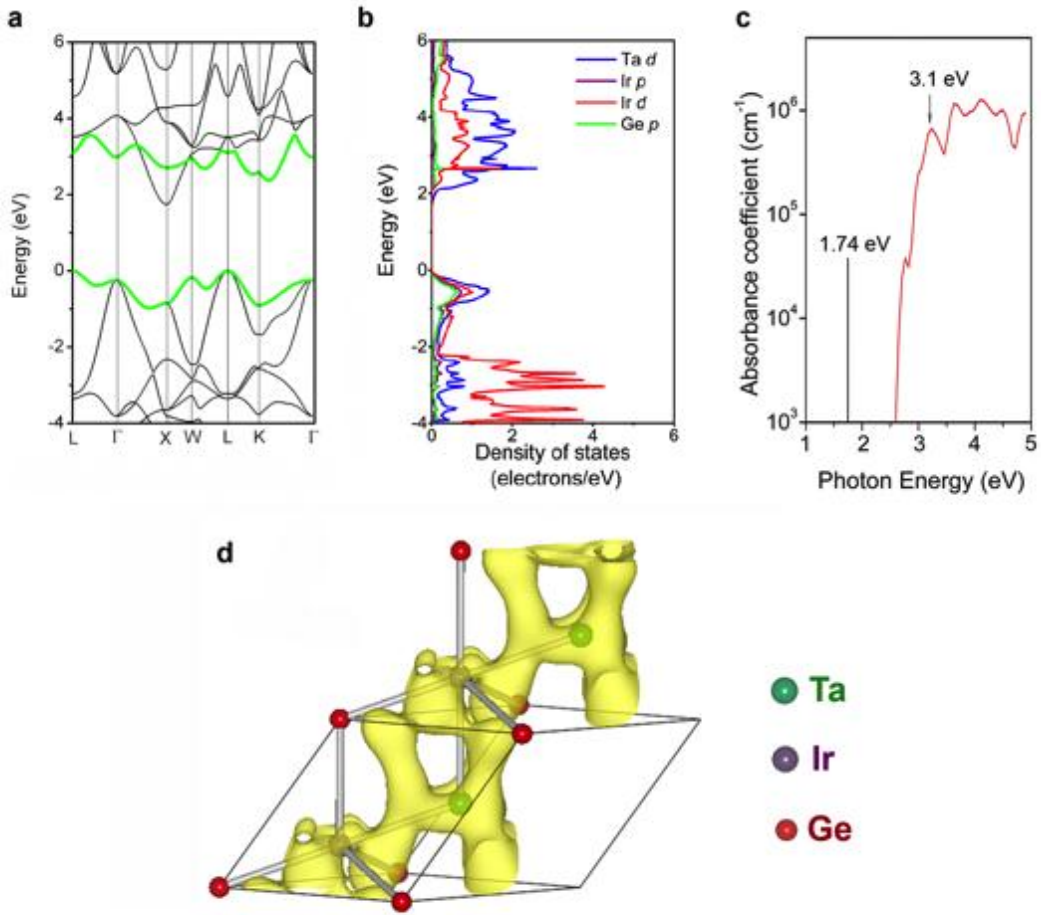

**Figure 2 | Electronic structure and optical properties of TaIrGe. a,** Band structure of TaIrGe. The two approximately parallel valence and conduction bands are highlighted by green giving a gap around 3.1 eV. **b.** Angular momentum and site decomposed partial density of states in TaIrGe. **c,** Calculated optical absorption coefficient. The black vertical line indicates the predicted minimum band gap. **d,** Wavefunction square of the VBM state located at L valley in TaIrGe (isosurface at $|\psi|^2 = 0.213$).

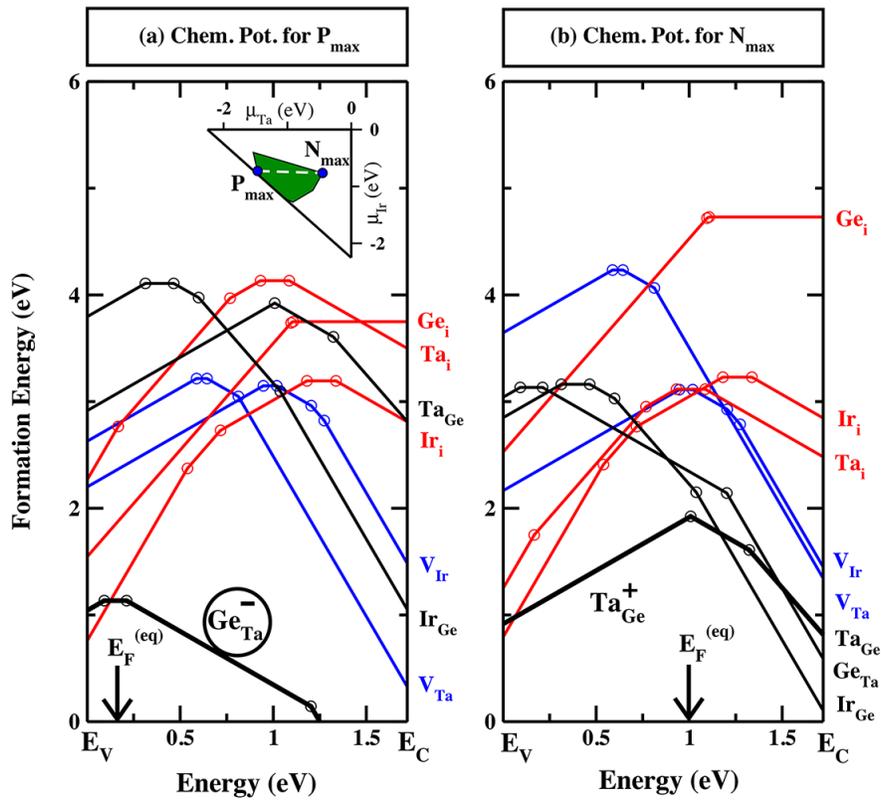

**Figure 3 | Formation enthalpies** of selected intrinsic defects of TaIrGe as a function of Fermi level $E_F$ at the chemical condition $(\mu_{Ta}, \mu_{Ir}, \mu_{Ge})$ that yields the highest p-type carrier concentration, $P_{max}$ (a) and that yields the highest n-type concentration, $N_{max}$ (b) within the stability triangle (inset of **a**).

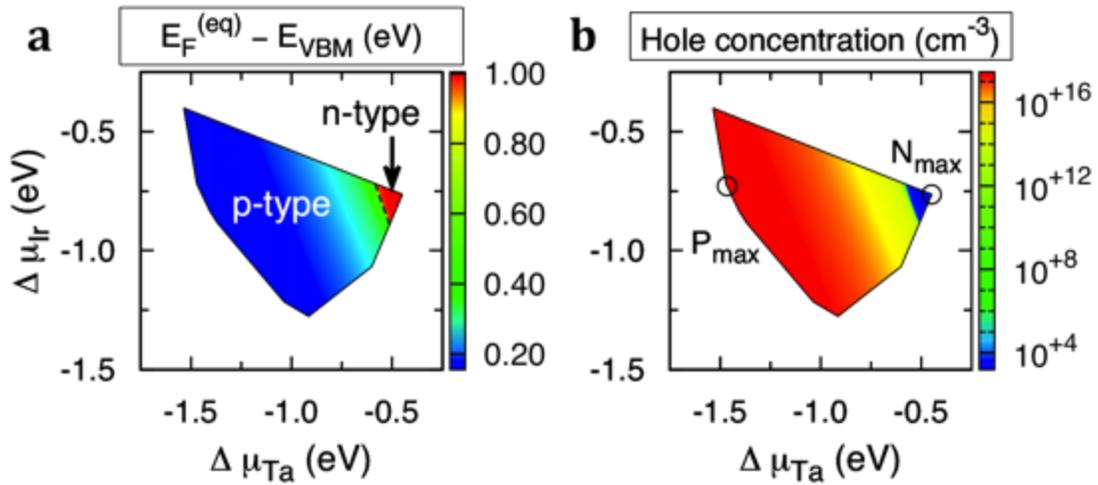

**Figure 4 | Equilibrium Fermi level (in eV) relative to VBM a** and hole concentration (in cm$^{-3}$) **b** due to intrinsic defects in the stable range of chemical potentials for TaIrGe. The simulation takes over an initial temperature of 1200 K, then quenches to room temperature. The chemical potential of Ge ($\Delta\mu_{Ge}$) used in defect calculation is determined by the formation enthalpy of TaIrGe $\Delta H_f$(TaIrGe) = −0.75 eV/atom, $\Delta\mu_{Ta}$ and $\Delta\mu_{Ir}$. The competing phases are shown in Supplementary Table S2.

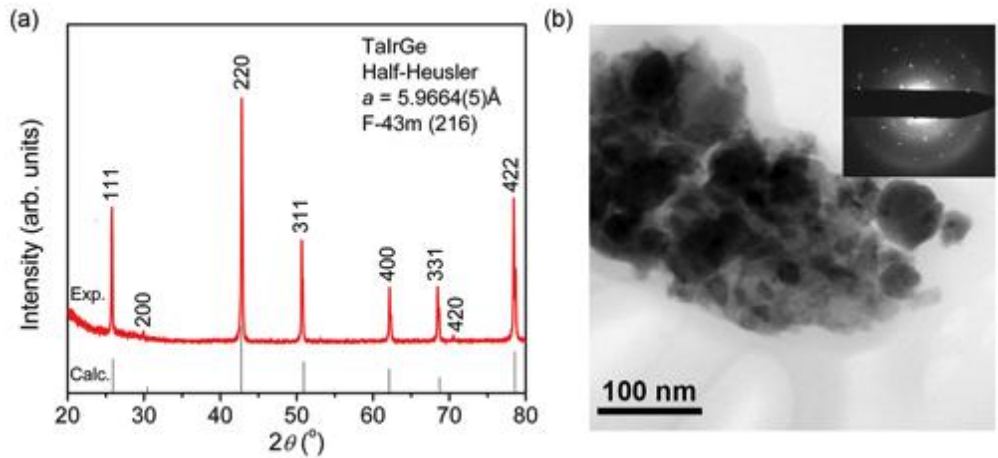

**Figure 5 | Measured crystal structure of TaIrGe. a,** Calculated and experimental XRD pattern of TaIrGe semiconductor. The final phase crystal structure, refined lattice constant, space group and space group number are shown. **b,** Representative bright field TEM image of TaIrGe. Inset shows a selected area electron diffraction (SAED) pattern along at a 110-electron-incidence displaying the single crystal nature of the individual grains.

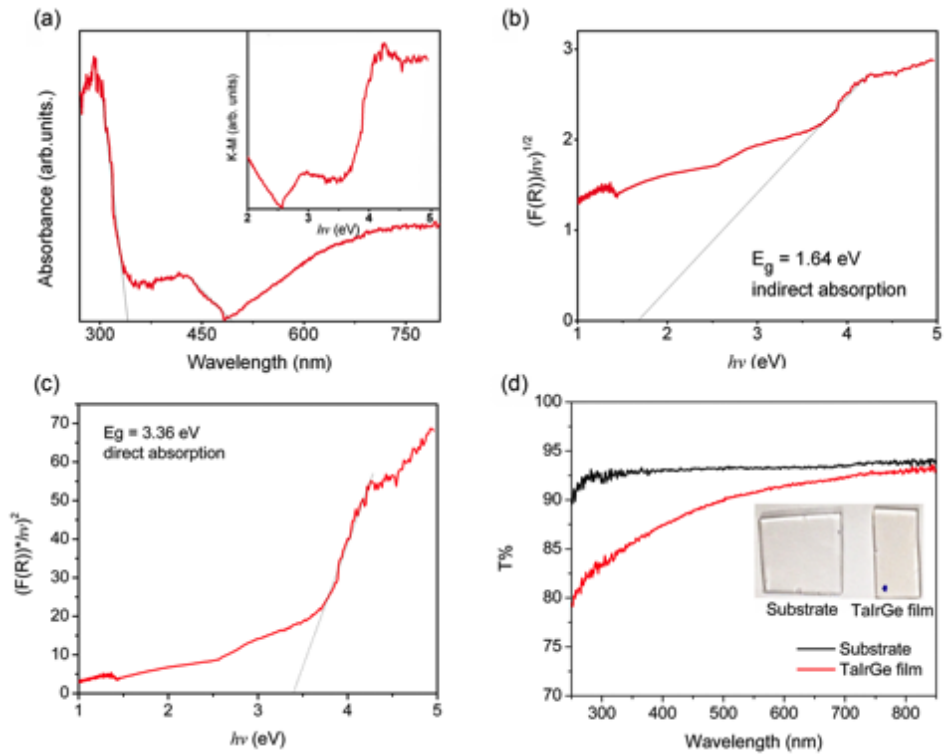

**Figure 6 | Measured optical property of TaIrGe semiconductor. a,** Optical spectra. The absorption spectra were obtained from the ultraviolet/visible diffuse reflectance spectra converted by the Kubelka-Munk[26] method (arbitrary units). **b, c,** The band gap of TaIrGe semiconductor determined by indirect and direct absorption, respectively. The optical band gap was determined by applying Tauc plots.[31] **d,** Transmittence for TaIrGe thin film (~30nm thick) grown on Quartz substrate using pulsed laser deposition (PLD) method. The images of the transparent quartz substrate and TaIrGe thin film on substrate are shown in the inset of **d**.